\documentstyle[preprint,aps]{revtex}

\begin{document}

\draft

\preprint{\begin{tabular}{l}
\hbox to\hsize{ \hfill BROWN-HET-1094}\\[-3mm]
\hbox to\hsize{ \hfill ANL-HEP-PR-97-64}\\[5mm] \end{tabular} }
\bigskip

\title
{Bounds on the lightest Higgs boson mass with three and four fermion generations}
\author{David Dooling, Kyungsik Kang and Sin Kyu Kang$^{\ast}$}

\address{Department of Physics\\
Brown University, Providence RI 02912, USA \\
$^{\ast}$ School of Physics \\
Korea Institute for Advanced Study \\
Seoul 130-012, Korea}


\maketitle

\begin{abstract}
We present lower bounds on the Higgs boson mass in the Standard Model with three and four fermion generations SM(3,4), as well as upper bounds on the lightest Higgs boson mass in the minimal supersymmetric extension of the SM with three and four generations MSSM(3,4).
Our analysis utilizes the SM(3,4) renormalization-group-improved one-loop effective potential of the Higgs boson to find the upper bounds on the Higgs mass in the MSSM(3,4) while the lower bounds in the SM(3,4) are derived from considerations of vacuum stability.
All the bounds increase as the degenerate fourth generation mass increases, providing more room in theory space that respects the increasing experimental lower limit of the Higgs mass.
\end{abstract}


\tightenlines


\centerline {\rm\bf I. Introduction}
With the discovery of the top quark at the Tevatron, the Higgs boson is now the only unknown sector of the standard model (SM).
Despite the remarkable agreement between theory and experiment exhibited in electroweak precision measurements, it is generally believed that the SM is not the final theory of elementary particle interactions.
The SM suffers from various theoretical blemishes, most notably the naturalness/fine-tuning problem associated with one fundamental physical Higgs scalar; i.e. the vacuum expectation value is quadratically unstable against radiative corrections. 
Another aesthetically unappealing feature of the SM is the lack of gauge unification at some GUT scale, typically $O(10^{16})$ GeV.
The minimal supersymmetric extension of the standard model (MSSM) addresses both of these issues \cite{1}. 
Because of the nature of supersymmetry (SUSY), the Higgs sector in the MSSM consists of two CP-conserving Higgs doublets with opposite hypercharge assignments \cite{2}.

 The quartic coupling of the Higgs potential is arbitrary in the SM and thus the Higgs boson mass is usually considered an adjustable parameter.
Nevertheless, if certain theoretical assumptions are imposed, upper and lower bounds on the Higgs boson mass can be obtained \cite{3}.
The requirement of vacuum stability yields a lower bound on the Higgs boson mass which depends on the top quark mass and the cut-off scale $\Lambda$ beyond which the SM is no longer valid, while an upper bound follows from requiring that no Landau singularity appears up to a scale $\Lambda$.
In the MSSM, intrinsic upper bounds on the lightest Higgs boson mass are obtained from the quartic higgs coupling which is no longer artibtrary but is constrained by SUSY \cite{2}.

Much previous work has been done in this area \cite{4,5,6,7,8}, but previous
determinations of the Higgs boson mass bounds in the MSSM and the SM have been forced to treat the top mass $(m_{t})$ as an unknown parameter.
 These results display the Higgs boson mass bounds as a function of the top mass for a given set of parameters.
With increasing precision of the measurement of the top quark mass, the aforementioned bounds and an eventual measurement of the Higgs mass can serve to either validate or discard both the MSSM and the SM, even though the couplings of the lightest MSSM Higgs boson are indistinguishable from those of the SM Higgs boson.
For a top quark mass $\lesssim 170$ GeV, the MSSM upper bound exceeds the SM lower bound.
If a Higgs mass lies between these two bounds for such a top quark mass, then no new information is obtained.
If the Higgs mass is measured with a value exceeding the MSSM upper bound and the SM lower bound, then the SM remains the only viable theory, while if the measurement lies below the MSSM upper bound and the SM lower bound, then only the MSSM survives.
The most interesting situation is realized for a top quark mass $<$ 175 GeV and when the Higgs mass lies in a region between the two bounds, with the SM lower bound exceeding the MSSM upper bound.
Such a situation demands the introduction of new physics.
For example, if we fix the running top quark mass to be 170 GeV, which corresponds to a top quark pole mass of $\sim$ 177 GeV, the possible scenarios are as follows.
For the supersymmetry breaking scale $M_{susy}$= 10 TeV and $\cos^{2}{2 \beta}=1$ and maximum threshold corrections, the upper bound on the physical Higgs mass is $\sim$ 142 GeV, while for the same set of parameters and $M_{susy}=$ 1 TeV, the upper bound is $\sim$ 135 GeV. 
The lower bound from the SM with the cut off scale $\Lambda=10^{19}$ GeV where new physics becomes absolutely necessary is $\sim$ 148 GeV.
As the experimental lower bound for the Higgs boson mass continues to rise, these theoretical bounds become more and more relevant.
If the MSSM is a faithful description of nature, then $M_{H}< 135(142)$ GeV for $M_{susy}= 1(10)$ TeV and so should be detected in the very near future.
The Higgs boson may be detected with $M_{H}<135$ GeV and all of the attractive features of the MSSM, e.g. gauge unification, a taming of the naturalness/fine-tuning problem, etc. remain intact.
On the other hand, suppose the experimental lower limit on $M_{H}$ is pushed above 142 GeV. 
Such a scenario rules out the MSSM as a viable theory.
If the Higgs mass is eventually measured and is greater than $\sim$ 148 GeV, then the SM provides a consistent and faithful description of nature all the way up to energy scales of $M_{planck}$.
For $M_{susy}=1$ TeV, the region 135 GeV$< M_{H} < 148$ GeV rules out both the SM and the MSSM.
If a Higgs boson is measured with a mass lying within this range consistent with all electroweak precision data to date, should we simply abandon the SM and all of its remarkable successes?
Noting that all the precision electroweak data indicates that $M_{H} <$ 370 GeV at one sigma order \cite{9}, we take the opposite approach, in the sense that we add an extra representation, a fourth generation, to both theories and calculate the Higgs mass bounds in the standard model with four generations (SM4) and in the minimal supersymmetric extension of the standard model with four generations (MSSM4).
This hypothetical fourth generation will differ from the familiar three in more ways than one, but most importantly the fourth generation neutral lepton must be massive with a mass greater than half the Z boson mass in order to not disturb the agreement 
between theory and experiment in measurements of the Z width \cite{10}.

So as best to be able to use the aforementioned bounds as a filter to discriminate between the MSSM Higgs boson and the pure SM Higgs boson, we will fix $\Lambda$ to be $10^{19}$ GeV, the scale where we know new physics becomes necessary in order to incorporate gravity.
It is well known that lower values of $\Lambda$ relax the SM lower bounds \cite{12}, but we note that the lower bounds on the SM Higgs boson mass are insensitive to the precise value of $\Lambda$ for large $\Lambda$; i.e. for $10^{11}$ GeV$ < \Lambda < 10^{19}$ GeV.


Several authors have derived the upper and lower bounds on the Higgs boson mass in the SM with extra generations as a function of the extra fermion masses \cite{11,12,13,14}.
 The bounds on the lightest Higgs boson mass in the MSSM4 have been calculated very recently in \cite{15}.
However, those bounds are based on one-loop order renormalization.
As pointed out by Casas et al. \cite{4}, the one-loop predictions of mass bounds of 
the SM Higgs and the lightest MSSM Higgs boson are strongly dependent on the scale at which the Higgs boson masses are calculated, while the two-loop results are almost flat as a function of scale.
Thus, two-loop results are more stable and meaningful.
In this paper, we would like to examine the physical (pole) Higgs boson mass bounds in the SM4 and the MSSM4 up to the two-loop level.

 The search for the Higgs boson being one of the major tasks along with that for supersymmetric sparticle and fourth generation fermions at future accelerators such as LEP200 and LHC makes it a theoretical priority to examine the bounds on the Higgs boson mass in the SM and its supersymmetric extension and to look for any distinctive features.
 The actual measurement of the Higgs boson mass could serve to exclude or at least to distinguish between the SM(3,4) and the MSSM(3,4) models for electroweak symmetry breaking.
 In order to carry out this procedure,
we adopt the basic assumptions that all superpartners of the SM particles as well as another Higgs scalar orthogonal to the lightest one have masses of the order of the supersymmetry breaking scale $M_{susy}$ so that the effective low-energy theory below 
$M_{susy}$ is equivalent to the SM with a single Higgs doublet (SM3 or SM4) \cite{16}.

Since the bounds on the Higgs boson mass in the models with four generations depend on the fourth generation fermion masses, it is convenient to impose some possible constraints on the fourth generation fermion masses.
The close agreement between the direct measurements of the top quark mass at the Tevatron and its indirect determination from the global fits of precision electroweak data including radiative corrections within the framework of the SM imply that there is 
no significant violation of the isospin symmetry for the extra generation \cite{12}.
Thus the masses of the fourth generation isopartners must be very close to degenerate; i.e.
\begin{displaymath}
\frac{\| m_{T}^{2}-m_{B}^{2}\|}{m_{Z}^{2}} \lesssim 1, \frac{\|m_{E}^{2}-m_{N}^{2}\|}{m_{Z}^{2}} \lesssim 1
\end{displaymath}
We are subscribing to the philosophy of trying to be as least destructive as possible to the SM in our proposal of an additional generation of fermions.
One argument often cited to disparage the SM as being an incomplete, effective theoy, a mere model, is the existence of the large number of free parameters.
For aesthetic reasons no different in principle from the accepted reasons for possibility of gauge coupling unification in the MSSM as additional support for its likely validity, we propose the simplest non-trivial solution to the phenomenologically imposed inequalities: a completely degenerate fourth generation, thus appealing to aesthetic proclivities as a compensation for extending the SM by this additional mass parameter.
One can think of the analogy of a forced, damped harmonic oscillator, where the more familiar three generations are analogous to the arbitrary initlal conditions contained in the transient solution, while the degenerate fourth generation is analogous to the in-phase, steady-state solution.
Recently, the limit on the masses of the extra neutral and charged leptons, $m_{N}$ and $m_{E}$, has been improved by LEP1.5 to $m_{N} > 59$ GeV and $m_{E} > 62$ GeV \cite{12}.
Our analysis yields upper bounds on $m_{4}$ that are the same in both the MSSM4 and SM4 cases.
As will be discussed in detail later,
the upper bound derived in the SM4 case results from requiring vacuum stability with a cut-off at the Planck scale $\Lambda=10^{19}$ GeV while that derived in the MSSM4 results from requiring pertubative validity out to the GUT scale.
Imposing these constraints leads us to restrict the range of $m_{4}$ to 50 GeV  $< m_{4} < 110$ GeV.
If we restrict $m_{4}$ to lie within this range of values, why has there been
no direct experimental observation of a fourth generation at the Tevatron?
As we mentioned earlier, our proposed fourth generation differs qualitatively from the other three.
Because of the mass degeneracy between isodoublets, W decay products resulting from a transition between fourth generation isopartners will be vanishingly soft and such a scenario could be hidden in the data.
Finally, we will assume that the fourth generation quarks do not mix with the known quarks. This assumption is plausible since the mixing angles are so small that the new  particles can leave the Tevatron detectors without decaying.

Our paper is organized as follows.
In Section I,
we first present our one-loop effective potential(EP) improved by the two-loop renormalization group equations (RGE) analysis. 
We note that, to the best of our knowledge, the present work is the first to present and to apply the SM4 two-loop RGE with a massive fourth generation neutral lepton.
Just as the two-loop corrections contribute significantly to the resulting bounds in the three generation case \cite{4}, we find that our two-loop results provide substantial contributions to previous one-loop results for the four generation theories previously calculated in \cite{14,15}.
 An expression for the Higgs boson mass is derived up to next-to-leading logarithm order. 
 In Section II our method for solving the RGE so as to evaluate the expressions for the mass is described.
We first evaluate this expression at the appropiate energy scale in order to make a meaningful comparison with our earlier one-loop results.
Next, following the example of Casas et al.\cite{4}, we shift all running masses to the physical pole masses.
 Upper bounds are obtained by imposing different boundary conditions on the Higgs self-coupling $\lambda$ which correspond to different values of
 supersymmetric mixing parameters, tan$\beta$, and
supersymmetry breaking scale $M_{susy}$. 
Numerical results are presented and a bound on $m_{4}$ derived from requirements of pertubative validity out to the GUT scale. We then turn in section III to our analysis of the vacuum stability bounds in the SM3 and the SM4.
 Another independent upper bound on $m_{4}$ is derived which is the same as that derived in the MSSM4 case. 
Finally, we compare all four scenarios and discuss the errors associated with our results arising from uncertainties in the strong coupling constant and the value of the top quark mass.
All of the relevant renormalization group equations  are presented in the Appendix.

\section{Effective Potential Approach }

Let us begin by making two reasonable assumptions in our analysis of the MSSM 
bounds on the lightest Higgs boson mass.
As mentioned above, we assume that all of the sparticles have masses $O(M_{susy}$) or greater and that of the two Higgs isodoublets of the MSSM, one linear combination is massive, also with a mass of $O(M_{susy}$) or greater, while the other linear
 combination, orthogonal to the first, has a mass of the order of weak-scale symmetry breaking.
 With these two assumptions, it is clear that below the supersymmetry breaking scale $M_{susy}$, the effective theory is the SM.
This fact enables us to use the SM effective potential for the Higgs boson
when we treat the lightest Higgs boson in the MSSM.

As shown in \cite{17}, in order to calculate the Higgs boson mass 
 up to the next-to-leading logarithm approximation,
we must consider the one-loop EP 
improved by two-loop RGE for the $\beta-$ and $\gamma-$ functions of the
running coupling constants, masses and the $\phi-$field for the Higgs boson.

The two-loop RGE improved one loop EP of the SM4 is given by
\begin{equation}
 V_{1} = V_{(0)} + V_{(1)} 
\end{equation}
where
\begin{eqnarray}
 V_{(0)} &=& -\frac{1}{2}m^{2}(t)\phi_{c}^{2}(t) + \frac{1}{24}\lambda(t)
              \phi_{c}^{4}(t) \\
V_{(1)} &=& \sum_{i=1}^{5} \left(-\frac{\kappa_{i}}{64\pi^{2}}\right)
            h_{i}^4(t)\phi_{c}^{4}(t)
            \left[\ln\frac{h_{i}^{2}(t)\zeta^{2}(t)}{2}-\frac{3}{2}\right] 
\end{eqnarray}
Here $\phi_c$ is the classical field corresponding to the physical
Higgs boson $\phi$, $i = (t,T,B,N,E)$
, $~\kappa_{i}=3$ for $i = (t,T,B)$ and $\kappa_{i}=1$ for $i = (N,E)$ and $h_{i}$ is the Yukawa coupling of the i$^{th}$ fermion to the Higgs field while $\zeta(t)$ is a quantity to be defined below related to the anamolous dimension of the Higgs field.
We have not included the contributions from the gauge boson and
would-be Goldstone bosons because their contributions are negligible compared to those from the heavy fermions \cite{6}.
Note that our definition of $\lambda$ differs by a factor of $\frac{1}{3}$ from most author's conventions. 
The above expression for the EP was obtained using the modified minimal substraction 
($\overline{MS}$) renormalization scheme,
and all of the RGE for the running couplings constants and masses
listed in the Appendix correspond 
to the choice of the Landau gauge. In the $\overline{MS}$ scheme, the EP does not satisfy the usual homogenous RGE unless an appropriate $\phi_{c}$ - independent term, which is the one-loop contribution to the cosmological constant, is added to the above 
expression.
When such an appropriate term is included, the EP satisfies the usual homogeneous RGE:
\begin{equation}
 \left( D-\gamma_{\phi}\phi_{c}\frac{\partial}{\partial\phi_{c}} \right)V(\phi_{c},X_{i},\mu)=0
\end{equation}
where
\begin{equation}
 D= \mu\frac{\partial}{\partial\mu}+ \beta_{X_{i}}\frac{\partial}
     {\partial X_{i}},
\end{equation}
\begin{eqnarray*}
 X_{i}= \left( \lambda, h_{t},h_{T},h_{B},h_{N},h_{E},m^{2},g_{1},g_{2},g_{3}\right),
\end{eqnarray*}
$\gamma_{\phi}$ is the anomalous dimension of the Higgs field.
All of the parameters $X_{i}$ run with energy, and so the RGE for the EP essentially states that a change in energy scale can be compensated for by a change in the values of the coupling constants and masses to ensure the renormalization group invariance of the EP. 
Thus, the EP can be written as
\begin{eqnarray}
 V(\phi_{c},X_{i},\mu) &=& V(\phi_{c}(t),X_{i}(t),\mu(t)) \nonumber \\
       &=& \Omega (X_i(t), \mu(t))+
  V_{0}(\phi_{c}(t),X_{i}(t))+V_{(1)}(\phi_{c}(t),X_{i}(t),\mu(t))
\end{eqnarray}
where $\Omega (X_{i}(t), \mu(t))$ is the $\phi_{c}$-independent piece alluded to above and knowledge of whose exact form is not necessary for our purposes.
 The $X_{i}$ are determined by the set of equations
\begin{equation}
 \frac{dX_{i}}{dt}= \beta_{X_{i}}(X_{j}(t)) 
\end{equation}
with the boundary conditions
\begin{equation}
 X_{i}(0)=X_{i},
\end{equation}
and
\begin{eqnarray}
\zeta(t) &=& \exp \left( -\int_{0}^{t} \gamma_{\phi}(t^{\prime})dt^{\prime}\right)
\\
\phi_{c}(t) &=& \phi_{c} \zeta(t)
\end{eqnarray}
and
\begin{equation}
\mu(t)=\mu\exp(t)
\end{equation}
where $\mu $ is a fixed scale.

In order to keep track of the consistency of our pertubation calculation, we expand the EP and the RGE for the running parameters in powers of $\hbar$, essentially an expansion in the number of loops:
\begin{eqnarray*}
V & = & \Omega + V_{(0)} + \hbar V_{(1)} \\
\beta_{X_{i}} & = & \hbar \beta_{X_{i}}^{(1)} + \hbar^{2} \beta_{X_{i}}^{(2)} \\
\gamma_{\phi}&=& \hbar \gamma_{\phi}^{(1)} + \hbar^{2}\gamma_{\phi}^{(2)} + \ldots
\end{eqnarray*}
So up to terms of $O(\hbar$), the RGE for the one-loop EP becomes
\begin{equation}
D \Omega+ \hbar \left( \beta_{X_{i}}^{(1)} \frac{\partial V_{(0)}}{\partial X_{i}} - \gamma_{\phi}^{(1)} \phi_{c}\frac{\partial V_{(0)}}{\partial \phi_{c}} \right) + \mu\frac{\partial}{\partial \mu} \left( \hbar V_{(1)}(\phi_{c}(t)) \right)=0
\end{equation}
Using Eq.(3) for $V_{(1)}(\phi_{c}(t))$,
this expression becomes:
\begin{equation}
D \Omega +\hbar \left\{ \beta_{X_{i}}^{(1)} \frac{\partial V_{(0)}}
{\partial X_{(i)}} - \gamma_{\phi}^{(1)}\phi_{c}\frac{\partial V_{(0)}}
{\partial \phi_{c}}+
\frac{1}{32\pi^{2}}\phi_{c}^{4} \left( \sum_{i=1}^{5} \kappa_{i} 
h_{i}^{4} \right) \right\}= 0 .
\end{equation}
We note that these expressions will be used later to eliminate terms  involving $D\Omega$.

Our first step is to simply evaluate $\partial V_{1}/\partial \phi_{c}$.
Using the chain rule and choosing $\phi_{c} = \mu(t)$, we have
\begin{eqnarray}
\frac{\partial X_{i}}{\partial \phi_{c}(t)} &=& \hbar \beta_{X_{i}}^{(1)}(t)\frac{1}{\phi_{c}(t)}+ O(\hbar^{2}) \\
\frac{\partial \mu(t)}{\partial \phi_{c}(t)} &=& \frac{\mu(t)}{\phi_{c}(t)} + O(\hbar)
\end{eqnarray}
Hence $\frac{\partial}{\partial \phi_{c}} \left[ \Omega+ V_{(0)}+V_{(1)} \right]$ can be expanded with the following equations:
\begin{eqnarray}
\frac{\partial \Omega}{\partial \phi_{c}(t)} &=& \frac{1}{\phi_{c}(t)} \left\{ \mu(t) \frac{\partial \Omega}{\partial \mu(t)}+ \hbar \beta_{X_{i}}^{(1)}(t) \frac{\partial \Omega}{\partial X_{i}(t)} \right\} + O(\hbar^{2})\\
\frac{\partial V_{(0)}}{\partial \phi_{c}(t)} &=& V_{(0)}^{\prime}(\phi_{c}(t))+ \hbar \frac{\beta_{X_{i}}^{(1)}(t)}{\phi_{c}(t)}\frac{\partial V_{(0)}(\phi_{c}(t))}{\partial X_{i}(t)} + O(\hbar^{2})\\
\frac{\partial V_{(1)}(\phi_{c}(t))}{\partial \phi_{c}(t)} &=& 
-\sum_{i=1}^{5} \frac{\hbar \kappa_{i} h_{i}^{4}}{16 \pi^{2}}\phi_{c}^{3}\left[ \ln \frac{h_{i}^{2} \zeta^{2}(t)}{2} -\frac{3}{2} \right] +O(\hbar^{2}).
\end{eqnarray}
We can also write the first derivative of the EP as
\begin{equation}
\frac{\partial V}{\partial \phi_{c}(t)}= -m^{2}\phi_{c}(t)+\frac{1}{6} \lambda(t) \phi_{c}^{3}(t)+\frac{\partial \Omega}{\partial \phi_{c}(t)} + \hbar \frac{\partial V_{(0)}}{\partial \phi_{c}(t)}.
\end{equation}
Because $\Omega$ has no explicit $\phi_{c}(t)$ dependence, we may use the previously obtained equations (12) and (13) to write
\begin{eqnarray*}
\mu \frac{\partial \Omega}{\partial \mu}+\hbar \beta_{X_{i}} \frac{\partial \Omega}{\partial X_{i}} = \hbar \left \{ \gamma_{\phi}^{(1)} \phi_{c} \frac{\partial V_{(0)}}{\partial \phi_{c}}-\beta_{X_{i}}^{(1)} \frac{\partial V_{(0)}}{\partial X_{i}} -\frac
{1}{32 \pi^{2}}\phi_{c}^{4} \left(\sum_{i=1}^{5} \kappa_{i} h_{i}^{4} \right) \right\}.
\end{eqnarray*}
It follows from the above equations (16,17,18,19) that
\begin{eqnarray}
\frac{\partial V}{\partial \phi_{c}(t)} &=& \left(1+\hbar \gamma_{\phi}^{(1)} \right) \left( -m^{2}\phi_{c}(t) + \frac{1}{6} \lambda(t) \phi_{c}^{3}(t) \right) \nonumber\\
& & + \hbar \left\{\sum_{i=1}^{5} -\frac{\kappa_{i}}{16 \pi^{2}} h_{i}^{4}\phi_{c}^{3}\left[ \ln \frac{h_{i}^{2} \zeta^{2}}{2} - 1 \right] \right\}.
\end{eqnarray}
Further differentiation of $\partial V/\partial \phi_c(t)$ by $\phi_c(t)$ gives
\begin{eqnarray}
 \frac{\partial^2 V}{\partial \phi^2_c(t)} &=&
 (1+\hbar \gamma^{(1)}_{\phi}(t))\left[-m^2(t)+\frac{1}{2}\lambda(t)\phi^2_c(t)
  \right]
+\hbar\left[-\beta^{(1)}_{m^2}(t)+\frac{1}{6}\beta^{(1)}_{\lambda}(t)\phi^2_c(t)\right]
\nonumber \\
 & & +\hbar \left\{\sum_{i=1}^{5} -\frac{3\kappa_{i}}{16 \pi^{2}} h_{i}^{4}\phi_{c}^{2}\left[ \ln \frac{h_{i}^{2} \zeta^{2}}{2} - 1 \right] \right\} 
\end{eqnarray}
Note that the three generation expression can be obtained by simply
dropping all the terms involving the fourth generation Yukawa couplings
in the above formalism.

We use the minimum condition $\partial V/ \partial \phi_{c}(t_{v}) = 0$ at $\phi_{c}(t_{v})=v$, and obtain for the lightest Higgs boson at the two-loop level
\begin{eqnarray}
m^2_{\phi} &=& \left( \frac{\partial^2 V}{\partial \phi^2_c(t)}\right)_{
    \phi_c(t_v)=v} \nonumber \\
&=& \frac{1}{3} \lambda(t_{v}) v^{2} +\hbar v^{2} \left\{ \frac{1}{6} \beta_{\lambda}^{(1)}(t_{v}) -\frac{\lambda}{6} \frac{\beta_{m^{2}}^{(1)}(t_{v})}{m^{2}} +\frac{1}{3} \lambda \gamma_{\phi}^{(1)} \right. \nonumber \\
& & + \left.\sum_{i=1}^{5}-\frac{\kappa_i}{8 \pi^{2}}h_{i}^{4}
     \left[ \ln \frac{h_{i}^{2} \zeta^{2}}{2}-1\right]\right\}
    + O(\hbar^2)
\end{eqnarray}

Substituting in the appropiate expressions for the beta functions given in the 
Appendix, we finally obtain the expression for the Higgs boson mass:
\begin{eqnarray}
m_{\phi}^{2} &=& \frac{1}{3}\lambda v^{2}+\frac{\hbar v^{2}}{16 \pi^{2}}\left\{ \frac{\lambda^{2}}{3} +2\lambda (h_{t}^{2}+h_{T}^{2}+h_{B}^{2}) +\frac{2\lambda}{3}(h_{E}^{2}+h_{N}^{2})\right. \nonumber \\
& & -\frac{\lambda}{2}(3 g_{2}^{2}+g_{1}^{2}) + \frac{9}{8}g_{2}^{4} +\frac{3}{4}g_{1}^{2}g_{2}^{2} + \frac{3}{8}g_{1}^{4} \nonumber \\
& &-6h_{t}^{4}\ln\frac{h_{t}^{2}\zeta^{2}}{2} -6h_{T}^{4}\ln\frac{h_{T}^{2}\zeta^{2}}{2}-6h_{B}^{4}\ln\frac{h_{B}^{2}\zeta^{2}}{2} \nonumber \\
& & \left. -2h_{E}^{4}\ln\frac{h_{E}^{2}\zeta^{2}}{2}-2h_{N}^{4}\ln\frac{h_{N}^{2}\zeta^{2}}{2} \right\} + O(\hbar^{2})
\end{eqnarray}
where the three generation case is obtained by simply letting the fourth generation Yukawa couplings go to zero.

Following the method similar of Kodaira et al. \cite{6}, we arrive at the appropiate energy scale at which to evaluate $\frac{\partial^{2} V}{\partial \phi_{c}(t)^{2}}$ by requiring $\frac{\partial V}{\partial \phi_{c}(t_{v})} =$ 0 at the scale $t_{v}$ where $\phi_{c}(t_{v}) = v = (\sqrt{2} G_{F})^{-\frac{1}{2}}$ = 246 GeV.
Using equations (9),(10), and (11), and the condition $\phi_{c}(t_{v})=v$, $t_{v}$ is found to satisfy
\begin{displaymath}
t_{v} = \ln \frac{v}{\mu} + \int_{0}^{t_{v}} \gamma_{\phi}(t^{\prime})dt^{\prime}.
\end{displaymath}
We then evaluate the first and second derivatives of the EP at the scale $t_{v}$ where $\phi_{c}(t_{v}) = v$.
In the above relation satisfied by $t_{v}$, $\mu$ is a fixed, constant mass scale.
In the MSSM theories, we will take $\mu$ to be $M_{susy}=1$ or 10 TeV, while the SM lower bounds will be derived after choosing $\mu= \Lambda= 10^{19}$ GeV.

 We define the running Higgs mass as:
\begin{displaymath}
m_{h}^{2}(t)= \frac{m_{h}^{2}(t_{v}) \zeta^{2}(t_{v})}{\zeta^{2}(t)}.
\end{displaymath}
The physical, pole masses are related to the running masses via the following equations:
\begin{displaymath}
M_{i}= \left[ 1 +\beta_{i} \frac{4 \alpha_{3}(M_{i})}{3 \pi} \right] m_{i}(M_{i})
\end{displaymath}
\begin{displaymath}
M_{H}^{2} = m_{h}^{2}(t) + Re\Pi(M_{H}^{2}) - Re\Pi(0)
\end{displaymath}
where $\beta_{i}$ = 0 for i = (N,E) and = 1 for i = (t,T,B).
$\Pi(q^{2})$ is the renormalized electroweak self-energy of the Higgs boson.
The expression used for $\Delta \Pi \equiv Re\Pi(p^{2}=M_{H}^{2})-Re\Pi(p^{2}=0)$ can be found in appendix D.
We follow exactly the derivation given in \cite{4}, but modify the fermion contribution appropriately to include the additional fermions with the appropriate color factors.

%


%
\section{Upper bound on $m_{H}$ in MSSM(3,4)}
In order to calculate the Higgs boson mass from Eq. (23), we solve 
the fully-coupled set of two-loop RGE for all of the couplings, the mass parameter and the anomalous dimension of the field $\phi$.

For the five Yukawa couplings, we impose the boundary conditions:
\begin{equation}
h_i(m_i)=\frac{\sqrt{2}m_{i}}{v}
\end{equation}
where $i = (t,T,B,N,E)$.
We choose the running top quark mass to be 170 GeV.
A more detailed estimation of the uncertainty to be assigned to our results 
arising from uncertainties in the top quark mass and in the strong coupling 
constant will be given later. 
For the gauge couplings we impose the boundary conditions at $M_{z}$ = 91.1884 GeV, \cite{10}:
\begin{eqnarray*}
\alpha_{3}(M_{z})&=&0.115\\
\alpha_{2}(M_{z})&=&0.0336\\
\alpha_{1}(M_{z})&=&0.0102\\
\end{eqnarray*}
The boundary condition for $\lambda$ defined at $M_{susy}$ is
\begin{equation}
\frac{\lambda}{3}(M_{susy})=\frac{1}{4}\left[g_{1}^{2}(M_{susy})+g_{2}^{2}(M_{susy})\right]\cos^{2}(2\beta)+\frac{\kappa_{i}h_{i}^{4}(M_{susy})}{16 \pi^{2}}\left( 2\frac{X_{i}}{M_{susy}^{2}}-\frac{X_{i}^{4}}{6 M_{susy}^{4}} \right)
\end{equation}
where $\kappa_{i}$ = 3 for $i = (t,T,B)$ and $\kappa_{i}$ = 1 for $i = (N,E)$ and $X_{i}$ is the supersymmetric mixing parameter for the ith fermion.
Because we are interested in bounds on the Higgs boson mass for given values of
$\tan\beta$ and $M_{susy}$, we only consider the cases of zero and 
maximum threshold corrections. 
Zero threshold corrections correspond to $X_{i}=0$.
Maximum threshold corrections occur for $X_{i}= 6 M_{susy}^{2}$. 
In this case the threshold correction becomes:
\begin{displaymath}
\Delta \lambda = \frac{9}{8 \pi^{2}}\left(3 h_{t}^{4}+3 h_{T}^{4}+3 h_{B}^{4}+ h_{E}^{4}+h_{N}^{4}\right)
\end{displaymath}
The appropiate three generation boundary conditions are simply obtained by once again setting all of the fourth generation Yukawa couplings to zero.
We use the Runge-Kutta method to numerically integrate the appropriate RGE.
 In order to derive the MSSM4 upper bound on the Higgs boson mass, we solve the system of coupled equations represented by equation (7), using the $\beta$ functions listed in appendix B, by demanding consistency between the initial values at the degenerate fourth generation mass scale and the boundary conditions for the total set of couplings.
 For the three generation case, we solve the same system using the $\beta$ functions listed in appendix A by demanding consistency between the initial values at the $M_{Z}$ scale and the boundary conditions for $\lambda$ and the top quark coupling.

Before presenting our MSSM(3,4) results, it is appropriate at this point to discuss our choice of the range of $m_{4}$.
Using the same criteria as Gunion et al. \cite{18}; i.e. demanding pertubative validity of the Yukawa couplings out to the GUT scale where the gauge coulings unify, we can determind an upper bound for $m_{4}$.
 This constraint translates into the condition 
$\frac{h_{i}^{2}}{4 \pi} \lesssim1$ and into the requirement that two-loop contributions
are small compared to the one-loop contributions in the RGE out to the GUT
scale.


 To check perturbative validity of the Yukawa couplings out to
the GUT scale, we solve the SM RGE subject to the boundary conditions given above and run the couplings out to the SUSY breaking scale, either 1 TeV or 10 TeV.
The MSSM4 beta functions listed in appendix C are then used to evolve the couplings out to the GUT scale.
The limit on the degenerate fourth generation fermion mass can
be extracted when $h_{N}$ just saturates the above criteria of pertubative validity at the GUT scale.
It is clear from the beta functions in appendix C that $h_{N}$ grows the fastest 
of all the Yukawa couplings when it is evolved  with the MSSM4 RGE 
because $\beta^{(1)}$ for $h_{N}$ does not have terms proportional to minus $g_{3}^{2}$.
Note that because the fourth generation neutral lepton carries less hypercharge
than does the fourth generation charged lepton, the terms proportional to 
minus the $g_{1}^{2}$ in its RGE are less significant for $h_{N}$ than for 
$h_{E}$, allowing the terms contributing positively to $\beta^{(1)}$ for $h_{N}$ to dominate more readily.
For $M_{susy}=1$ TeV and 10 TeV, we have observed that the upper bound
on a degenerate fourth generation fermion mass $m_4 \simeq 110$ GeV.


Our numerical value for the upper bound of $m_4$ differ from those of Gunion et al. \cite{18} for at least two reasons.
First, while Gunion et al. only assume degeneracy among the quark and lepton sectors separately, we adopt a completely degenerate fourth generation, principally so as to reduce the number of parameters. 
Second, and more importantly, the discrepancy between the two results arises from different choices of $M_{susy}$ and application of appropriate RGE.
 Gunion et al. choose $M_{susy}$ as  $M_Z$ and
use the SUSY RGE even below the low SUSY breaking scale of $M_Z$.
 Their results allow for much heavier quarks $m_T\lesssim 178$ GeV,
$m_B\lesssim 156$ GeV, while they restrict the leptons to lower masses
$m_N, m_E\lesssim 86$ GeV.
As mentioned above, we choose $M_{susy}$ as 1 TeV or 10 TeV and use the SM4 RGE to evolve couplings below $M_{susy}$.

Even with this derived bound on a possible fourth generation degenerate fermion
mass, we continue to consider the cases of increasing $m_{4}$ all the way out 
to $m_{4}$ = 250 GeV, just as in our one-loop results \cite{15}.
These ``out of bounds'' cases are still considered because neglecting them 
involves the implicit assumption that no new physics becomes manifest between 
$M_{susy}$ and the GUT scale.
If this assumption is relaxed, the above requirement of pertubative validity 
all the way out to the GUT scale is clearly not necessary, it is only necessary
for pertubative validity to hold out to some intervening mass scale between 
$M_{susy}$ and $M_{GUT}$.
The choice of 250 GeV as the cut-off for $m_{4}$ arises from requiring 
$h_{i}^{2}/4 \pi \lesssim 1$ out to $M_{susy}$. 
It is important to note that for $m_{4} > 110$ GeV, the possibility of gauge unification is lost and the hierarchy problem becomes less drastic, thereby weakening the need for supersymmetry.
As mentioned in the introduction, one of the most appealing features of the MSSM theories is the possibility of gauge unification. 
All values of $m_{4} >$ 110 GeV eradicate this feature.
Another reason for considering supersymmetry is that it helps provide a solution to the hierarchy problem.
Supersymmetry makes sense out of the tremendous difference between the weak scale and the GUT scale where new physics holds.
As $m_{4}$ increases, pertubative validity requires that new physics start at some scale $\Lambda$ less than the GUT scale.
$\Lambda$ descreases with increasing $m_{4}$.
As the difference between the weak scale and $\Lambda$ becomes smaller, the hierarchy problem becomes less and less drastic, thereby weakening the motivation for supersymmetry.

In Fig. (1) we present our numerical two-loop results for the lightest Higgs boson mass bounds in the MSSM4(solid lines) as a function of $m_{4}$ for the case $M_{susy}=1$ TeV and (a) no threshold corrections and (b) maximum threshold corrections.
In Fig. (1) and Fig. (2), the upper curve corresponds to $\cos^{2}(2 \beta)=1$ and the lower curve to $\cos^{2}(2\beta)=0$.
For the sake of comparison, we also present the lightest Higgs boson mass bounds in MSSM3(dotted lines) for the same paarameter sets.
Our three generation results are in agreement with those of Casas et al. \cite{4}.
Fig. 2 shows the case for $M_{susy}=10$ TeV.
In the analysis, we fix the running top quark mass $m_{t}(m_{t})=170$ GeV and $\alpha_{s}(M_{z}) = 0.115$.
After shifting to pole masses, we find $M_{t}$ = 177 GeV in all cases.
As $m_{4}$ increases $M_{H}$ of the MSSM4 increases rapidly.
Choosing $\cos^{2}(2 \beta)=1$ results in a greater upper bound on the Higgs mass, primarily because this choice requires a larger initial value of $\lambda(m_{4})$ which in turn makes $\lambda$ larger at the scale where the mass is evaluated.
As one can see from Figs. 1 and 2, the presence of the fourth generation raises the lightest Higgs mass compared to the three generation case.
In particular, in the range 50 GeV$ <  m_{4} <$ 110 GeV which permits a SUSY GUT, the difference between the MSSM3 and MSSM4 bounds on $M_{H}$ ranges from 7 GeV ($m_{4}$ = 50 GeV) to 30 GeV ($m_{4}=110$ GeV).
For the case $M_{susy}=10$ TeV, the difference becomes more pronounced.

In Figs. 3 and 4, we present the one-loop running Higgs mass $m_{h}$ (solid lines) and the two-loop running Higgs mass $m_{h}$ (dotted line) results in the MSSM4.
Our one-loop results are based on the analysis given in \cite{15}.
As can be seen from those figures, the difference between one-loop $m_{h}$ and two-loop $m_{h}$ results in the $M_{susy}=1$ TeV case becomes larger as $m_{4}$ increases, but the difference between the one and two-loop depends strongly on $\cos^{2}(2 \beta)$.
For $\cos^{2}(2 \beta)=1$, the two-loop results are positive with respect to the one-loop results, while for $\cos^{2}(2 \beta)=0$, the two-loop affects lower the one-loop results in the region 50 GeV $\lesssim m_{4} \lesssim$ 100 GeV.
A similar situation holds for the $M_{susy}=10$ TeV case, where $\cos^{2}(2 \beta)=0$ leads to negative two-loop contributions for $m_{4} \lesssim 80$ GeV.

The abscissa of our plots is to be interpreted as the running mass of the degenerate fourth generation leptons.
The pole masses of the fourth generation quarks, $M_{T}$ and $M_{B}$, are slightly greater than $m_{4}$, and this difference increases with increasing $m_{4}$.
For example, when $m_{4}$ = 50 GeV, both $M_{T}$ and $M_{B}$ are $\sim$ 52.5 GeV.
When $m_{4}$ = 170 GeV, both $M_{T}$ and $M_{B}$ are $\sim$ 177 GeV.
Lastly, when $m_{4}$ = 250 GeV, $M_{T}$ and $M_{B}$ are $\sim$ 261 GeV.

\section{LOWER BOUNDS ON $M_{H}$ IN THE SM}

In this section, we study the lower bounds on the Higgs boson mass in the SM with three and four fermion generations.
Let us describe our procedure for determining a lower bound on the Higgs boson mass in the SM.
Again, we employ a consistent two-loop analysis and use equation (23).
We first alert the reader to our phenomenologically viable assumption that the physical vacuum corresponding to the vaccum expectation of the Higgs field of 246 GeV is indeed the true vacuum of the effective potential; i.e. that the physical vacuum corresponds to a global, not merely a local, minimum of the effective potential.
This assumption is consistent with our intention to accept the SM as a truly valid theory and compute the consequences; i.e. to zeroth order there is no motivation to consider the physical vacuum to be anything other than the true vacuum.
If one considers the possibility that the physical vacuum is a meta-stable vacuum with a lifetime longer than the age of the universe, that there exist deeper minima of the potential, then the SM lower bounds on the Higgs boson mass become less stringent in general for certain choices of $\Lambda$ and $M_{top}$ \cite{21}.
But for $M_{top} \sim 177$ GeV and $\lambda = 10^{19}$ GeV, the SM3 absolute stability lower bound is relaxed by only $\sim O(5)$ GeV when one only immposes metastability requirements, and this small effect only becomes diminished with the inclusion of a fourth generation. 
We obtain lower limits on the SM Higgs boson mass by requiring stability of this observed vacuum \cite{3,19}.

Working with the two-loop RGE requires the imposition of one-loop boundary conditions on the running parameters.
As pointed out by Casas et al. \cite{5}, the necessary condition for vacuum stability is derived from requiring that the effective coupling $\tilde{\lambda}(\mu)>$ 0  rather than $\lambda > 0$ for $\mu(t) < \Lambda$, where $\Lambda$ is the cut-off beyond which the SM is no longer valid. 
The effective coupling $\tilde{\lambda}$ in the SM4 is defined as:
\begin{displaymath}
\tilde{\lambda}=\frac{\lambda}{3} -\frac{1}{16 \pi^{2}}\left\{ \sum_{i=1}^{5} 2 \kappa_{i} h_{i}^{4} \left[ \ln \frac{h_{i}^{2}}{2} - 1 \right] \right\}
\end{displaymath}
where the three generation case is simply the same as the above expression without the fourth generation Yukawa coupling contributions.
We then solve the RGE in the same manner as in the SUSY case but impose this boundary condition on $\tilde{\lambda}$.
Again, we fix $m_{t}(m_{t})$ = 170 GeV and $\alpha_{s}(M_{z})=0.115$.

Before discussing the Higgs boson mass bound, let us consider an upper bound on $m_{4}$.
This upper bound can be determined by considering the so-called ``triviality'' bound on the Higgs mass derived from the requirement that no Landau pole occurs for $\tilde{\lambda}$ all the way out to the cut-off scale $\Lambda$ \cite{20}.
If one assumes that $m_{4}$ could continue out to any value up to the cut-off $\Lambda$, then one would presumably obtain both a lower bound on the Higgs mass from considerations of vacuum stability as well as an upper bound from the requirement $\tilde{\lambda}<\infty$; i.e., that there be no Landau pole for $\tilde{\lambda}$ out to $\Lambda$.
But clearly $m_{4}$ must be bounded from above somehow, for otherwise all of the fourth generation Yukawa couplings will become nonpertubative at some scale well below $\Lambda$, which in turn would feed the other couplings via the RGE until they also become nonpertubative.
In this way, we can obtain an upper bound for $m_{4}$, the value of $m_{4}$ where the vacuum stability and ``triviality'' bounds coincide.
For $m_{t}(m_{t})=170$ GeV and $\Lambda = 10^{19}$ GeV, it turns out that the upper bound on $m_{4}$ is $\sim$ 100 GeV.
Interestingly, this value is close to the upper bound on $m_{4}$ derived in the MSSM4.

In Fig. 5(a), the upper curve is the SM4 lower bound on the Higgs mass derived from vacuum stability requirements.
The cut-off scale $\Lambda$ is $10^{19}$ GeV and $M_{susy}=10$ TeV.
The horizontal line just beneath it is the SM3 lower bound.
The curve starting below the SM3 lower bound is the MSSM4 upper bound with SUSY parameters $M_{susy}=10$ TeV, $\cos^{2}(2 \beta)=1$ and maximum threshold corrections; i.e., the absolute maximum MSSM4 upper bound.
The bottom flat line is the MSSM3 upper bound for the same set of SUSY parameters.
Fig. 5(b) displays the same results for $M_{susy}$ = 1 TeV and the same set of remaining SUSY parameters as above.
These curves intersect and divide the ($M_{H},m_{4}$) plane into several regions.
Eventual measurement of the Higgs mass and any experimental evidence for a fourth generation of fermions could serve as a filter to select among or to discard some of the four theories considered.
For example, a light Higgs mass ($<148$ GeV) would rule out both the SM3 and the SM4 independently of any considerations of $m_{4}$.
The region above the bottom curve completely rules out the MSSM3 and MSSM4. 
Considering just the four gerneration theories, it is interesting to note that the region between the two highest lying curves, where both the SM4 and the MSSM4 are excluded as viable theories, is increasing in area as $m_{4}$ increases.
This behaviour further suggests that any possible fourth generation fermions must have masses not much greater than 100 GeV.

\section{CONCLUDING REMARKS}

Recall that the results we present are all derived with a running top quark mass $m_{t}$ = 170 GeV and $\alpha_{3}(M_{z})=0.115$.
This $m_{t}$ results in all cases in an $M_{t}$ = 177 GeV.
We now discuss the uncertainties to be assigned to our results induced by the uncertainties in $M_{t}$ and $\alpha_{3}(M_{z})$ for all four theories.
For the MSSM3, $\Delta m_{t} = \pm5$ GeV induces shifts in $M_{t}$ of $\pm5$ GeV and in the Higgs boson mass upper bound of $\pm3$ GeV.
When we let $\alpha_{3}(M_{z})=.12$, the bounds shift downward by$ <$ 1 GeV and when $\alpha_{3}(M_{z})=.11$, the bounds shift upward by$ <$ 1 GeV.
The above results correspond to $M_{susy}$ = 1 TeV.
Choosing $M_{susy}=10$ TeV leaves the $\alpha_{3}(M_{z})$ behaviour unaltered, but the uncertainties in the Higgs mass upper bound induced by $\Delta m_{t} = \pm 5$ GeV grow to $\pm$ 5 GeV.
The MSSM4 uncertainties are virtually identical to the MSSM3 uncertainties.
For the SM3, letting $\Delta m_{t}$ = $\pm$ 5 GeV results in changes in the Higgs boson mass vacuum stability lower bound of $\pm$ 9 GeV.
For the SM4, the induced shift is similar, $\pm 10$ GeV.
For both the SM3 and the SM4 the uncertainties in $\alpha_{3}(M_{z})$ result in larger shifts of the Higgs boson mass than in the MSSM theories.
When $\alpha_{3}(M_{z})$ changes by $\pm$ .005, the lower bounds in the SM theories change by $\mp$ 6 GeV.

In conclusion, we have studied the upper bounds on the lightest Higgs boson mass $M_{H}$ in the MSSM with four generations by solving the two-loop RGEs and using the one-loop EP.
We have considered the fourth generation of quarks and leptons with the degenerate mass $m_{4}$.
Our implementation of the appropriate two-loop RGE makes our results the most precise determination of the SM4 and MSSM4 Higgs mass bounds to date.
We have shown how the bounds on $M_{H}$ depend on $m_{4}$.
 In the region of large $m_{4}$, the bounds on $M_{H}$ increase with increasing $m_{4}$, while they are relatively insensitive to $m_{4}$ in the region of small $m_{4}$.
An upper bound on $m_{4}$ (100 GeV$ < m_{4max} <$ 110 GeV) is derived in two independent ways; one from requiring pertubative validity all the way out to the GUT scale and one from requiring vacuum stability in the pure SM4 out to the Planck scale.
Finally, we have presented results for a lower bound on the Higgs boson mass in the SM4 arising from the same requirement of vacuum stability out to the Planck scale, the first two-loop results of its kind.

\section{ACKNOWLEDGEMENTS}

One of us (S.K.K.) would like to acknowledge KOSEF for a Post-doctoral Fellowship and also to thank the High-Energy Theory Group, Brown University for kind hospitality and support.
He would also like to thank D.W. McKay for helpful comments.
Two of us (K.K. and D.D.) would like to thank the ANL high-energy group for hospitality during the completion of this work and support provided in part by the U.S. Dept. of Energy, HEP DIV., under COntract W-31-109-Eng-38. 
We also wish to thank M. Machacek and M. Vaughn for helpful discussions concerning the RGE used in this investigation.
Support for this work was provided in part by U.S. Dept. of Energy Contract DE-FG-02-91er40688-Task A.

\newpage

\section*{Appendix}
\subsection{SM3}

In this appendix we present the RGE for the standard model with three and four generations. We also present the relevant MSSM4 RGE used to determine the bounds on $m_{4}$.
We define the constant A as $A \equiv 16 \pi^{2}$.
\begin{displaymath}
A\beta_{\lambda}^{(1)}=4 \lambda^{2} + 12\lambda h_{t}^{2} - 36h_{t}^{4} -3\lambda(3g_{2}^{2}+g_{1}^{2})+\frac{27}{4}g_{2}^{4} +\frac{9}{2}g_{2}^{2}g_{1}^{2}+\frac{9}{4}g_{1}^{4}
\end{displaymath}
\begin{displaymath}
A^{2}\beta_{\lambda}^{(2)}=-\frac{26}{3}\lambda^{3} - 24\lambda^{2}h_{t}^{2} + 6 \lambda^{2}(3 g_{2}^{2} + g_{1}^{2}) + \lambda \left\{ -3 h_{t}^{4} +h_{t}^{2}(80 g_{3}^{2}+\frac{45}{2}g_{2}^{2}+\frac{85}{6}g_{1}^{2})-\frac{73}{8}g_{2}^{4} + \frac{39}{4}g_{2}^{2}g_{1}^{2}+\frac{629}{24}g_{1}^{4} \right\}
\end{displaymath}
\begin{displaymath}
+180h_{t}^{6}-h_{t}^{4}(192g_{3}^{2}+16g_{1}^{2})+h_{t}^{2}(-\frac{27}{2}g_{2}^{4}+63g_{2}^{2}g_{1}^{2}-\frac{57}{2}g_{1}^{4}) +\frac{915}{8}g_{2}^{6}-\frac{289}{8}g_{2}^{4}g_{1}^{2}-\frac{559}{8}g_{2}^{2}g_{1}^{4}-\frac{379}{8}g_{1}^{6}
\end{displaymath}
\begin{displaymath}
A\beta_{h_{t}}^{(1)}= \frac{9}{2}h_{t}^{3} -h_{t}(8g_{3}^{2}+\frac{9}{4}g_{2}^{2}+\frac{17}{12}g_{1}^{2})
\end{displaymath}
\begin{displaymath}
A^{2}\beta_{h_{t}}^{(2)}=h_{t}\left\{-12 h_{t}^{4}-2\lambda h_{t}^{2} +h_{t}^{2}(36 g_{3}^{2}+ \frac{225}{16}g_{2}^{2}+\frac{131}{6}g_{1}^{2}) \right\}
\end{displaymath}
\begin{displaymath}
+h_{t}\left\{ \frac{1}{6}\lambda^{2}-108g_{3}^{4}+9g_{3}^{2}g_{2}^{2}+\frac{19}{9}g_{3}^{2}g_{1}^{2}-\frac{23}{4}g_{2}^{4}-\frac{3}{4}g_{2}^{2}g_{1}^{2}+\frac{1187}{216}g_{1}^{4}\right\}
\end{displaymath}
\begin{displaymath}
A\beta_{g_{3}}^{(1)}=-7g_{3}^{3}
\end{displaymath}
\begin{displaymath}
A^{2}\beta_{g_{3}}^{(2)}=g_{3}^{3}(-2h_{t}^{2}-26g_{3}^{2}+\frac{9}{2}g_{2}^{2}+\frac{11}{6}g_{1}^{2})
\end{displaymath}
\begin{displaymath}
A\beta_{g_{2}}^{(1)}=-\frac{19}{6}g_{2}^{3}
\end{displaymath}
\begin{displaymath}
A^{2}\beta_{g_{2}}^{(2)}=g_{2}^{3}(-\frac{3}{2}h_{t}^{2}+12g_{3}^{2}+\frac{35}{6}g_{2}^{2}+\frac{3}{2}g_{1}^{2})
\end{displaymath}
\begin{displaymath}
A\beta_{g_{1}}^{(1)}=\frac{41}{6}g_{1}^{3}
\end{displaymath}
\begin{displaymath}
A^{2}\beta_{g_{1}}^{(2)}=g_{1}^{3}(-\frac{17}{6}h_{t}^{2}+\frac{44}{3}g_{3}^{2}+\frac{9}{2}g_{2}^{2}+\frac{199}{18}g_{1}^{2})
\end{displaymath}
\begin{displaymath}
A\beta_{m^{2}}^{(1)}=m^{2}(2\lambda+6h_{t}^{2}-\frac{9}{2}g_{2}^{2}-\frac{3}{2}g_{1}^{2})
\end{displaymath}
\begin{displaymath}
A\gamma_{\phi}^{(1)}=3h_{t}^{2}-\frac{9}{4}g_{2}^{2}-\frac{3}{4}g_{1}^{2}
\end{displaymath}
\begin{displaymath}
A^{2}\beta_{\phi}^{(2)}=\frac{1}{6}\lambda^{2}-\frac{27}{4}h_{t}^{4}+h_{t}^{2}(20 g_{3}^{2}+\frac{45}{8}g_{2}^{2}+\frac{85}{24}g_{1}^{2})-\frac{271}{32}g_{2}^{4}-\frac{9}{16}g_{2}^{2}g_{1}^{2}+\frac{431}{96}g_{1}^{4}
\end{displaymath}
\subsection{SM4}
\begin{displaymath}
A\beta_{\lambda}^{(1)}=4\lambda^{2}+4\lambda\left[3h_{t}^{2}+3(h_{T}^{2}+h_{B}^{2})+(h_{E}^{2}+h_{N}^{2})\right]-12\left[3h_{t}^{4}+3(h_{T}^{4}+h_{B}^{4})+(h_{E}^{4}+h_{N}^{4})\right]
\end{displaymath}
\begin{displaymath}
-3\lambda(g_{1}^{2}+3g_{2}^{2})+\frac{9}{4}g_{1}^{4}+\frac{9}{2}g_{1}^{2}g_{2}^{2}+\frac{27}{4}g_{2}^{4}
\end{displaymath}
\begin{displaymath}
A^{2}\beta_{\lambda}^{(2)}=-\frac{25}{3}\lambda^{3}+6(g_{1}^{2}+3g_{2}^{2})\lambda^{2}-\lambda\left(\frac{2861}{36}g_{1}^{4}-\frac{117}{12}g_{1}^{2}g_{2}^{2}-\frac{7}{8}g_{2}^{4}\right)+\frac{241}{8}g_{2}^{6}-\frac{353}{24}g_{1}^{2}g_{2}^{4}
\end{displaymath}
\begin{displaymath}
-\frac{5991}{216}g_{1}^{4}g_{2}^{2}-\frac{4371}{216}g_{1}^{6}-192g_{3}^{2}(h_{t}^{4}+h_{T}^{4}+h_{B}^{4})-8g_{1}^{2}(2h_{t}^{4}+2h_{T}^{4}-h_{B}^{4}+3h_{E}^{4})
\end{displaymath}
\begin{displaymath}
-\frac{9}{2}g_{2}^{4}\left[(\frac{17}{12}g_{1}^{2}+\frac{9}{4}g_{2}^{2}+8g_{3}^{2})(h_{t}^{2}+h_{T}^{2})+(\frac{5}{12}g_{1}^{2}+\frac{9}{4}g_{2}^{2}+8g_{3}^{2})h_{B}^{2}+(\frac{5}{4}g_{1}^{2}+\frac{3}{4}g_{2}^{2})h_{E}^{2}+(\frac{1}{4}g_{1}^{2}+\frac{3}{4}g_{2}^{2})h_{N}^{2}\right]
\end{displaymath}
\begin{displaymath}
10\lambda\left[(\frac{17}{12}g_{1}^{2}+\frac{9}{4}g_{2}^{2}+8g_{3}^{2})(h_{t}^{2}+h_{T}^{2})+(\frac{5}{12}g_{1}^{2}+\frac{9}{4}g_{2}^{2}+8g_{3}^{2})h_{B}^{2}+(\frac{5}{4}g_{1}^{2}+\frac{3}{4}g_{2}^{2})h_{E}^{2}+(\frac{1}{4}g_{1}^{2}+\frac{3}{4}g_{2}^{2})h_{N}^{2}\right]
\end{displaymath}
\begin{displaymath}
+3g_{1}^{2}\left[( -\frac{57}{6}g_{1}^{2}+21g_{2}^{2})(h_{t}^{2}+h_{T}^{2})+(\frac{5}{2}g_{1}^{2}+9g_{2}^{2})h_{B}^{2}+(-\frac{25}{2}g_{1}^{2}+11g_{2}^{2})h_{E}^{2}+(\frac{5}{2}g_{1}^{2}+11g_{2}^{2})h_{N}^{2}\right]
\end{displaymath}
\begin{displaymath}
-8\lambda^{2}\left( 3h_{t}^{2}+3h_{T}^{2}+3h_{B}^{2}+h_{E}^{2}+h_{N}^{2} \right)-\lambda \left[ 3h_{t}^{4}+3h_{T}^{4}+3h_{B}^{4}+h_{E}^{4}+h_{N}^{4}\right]+6\lambda h_{T}^{2}h_{B}^{2}+2\lambda h_{E}^{2} h_{N}^{2}
\end{displaymath}
\begin{displaymath}
60\lambda\left[ 3h_{t}^{6}+3h_{T}^{6}+3h_{B}^{6}+h_{E}^{6}+h_{N}^{6}\right] -36(h_{T}^{4}h_{B}^{2}+h_{B}^{4}h_{E}^{2})-12(h_{E}^{4}h_{N}^{2}+h_{N}^{4}h_{E}^{2})
\end{displaymath}
\begin{displaymath}
A \beta_{h_{t}}^{(1)}=h_{t}\left[\frac{3}{2}h_{t}^{2}+3(h_{t}^{2}+h_{T}^{2}+h_{B}^{2})+h_{N}^{2}+h_{E}^{2}-\frac{17}{12}g_{1}^{2}-\frac{9}{4}g_{2}^{2}-8g_{3}^{2}\right]
\end{displaymath}
\begin{displaymath}
A^{2}\beta_{h_{t}}^{(2)}=h_{t}\left\{\frac{3}{2}h_{t}^{4}-\frac{9}{4}h_{t}^{2}\left(3(h_{t}^{2}+h_{T}^{2}+h_{B}^{2})+h_{E}^{2}+h_{N}^{2}\right)\right\}
\end{displaymath}
\begin{displaymath}
+  h_{t}\left\{         -\frac{9}{4}\left( 3(h_{t}^{4}+h_{T}^{4}+h_{B}^{4})+h_{E}^{4}+h_{N}^{4}-\frac{2}{3}h_{T}^{2}h_{B}^{2}-\frac{2}{9}h_{N}^{2}h_{E}^{2}\right) \right\}
\end{displaymath}
\begin{displaymath}
+h_{t}\left\{ \frac{\lambda^{2}}{6}-2\lambda h_{t}^{2} +\left(\frac{223}{48}g_{1}^{2}+\frac{135}{16}g_{2}^{2}+16g_{3}^{2}\right)h_{t}^{2}\right\}
\end{displaymath}
\begin{displaymath}
+h_{t}\left\{ \frac{5}{2} \left[ (\frac{17}{12}g_{1}^{2}+\frac{9}{4}g_{2}^{2}+8g_{3}^{2})(h_{t}^{2}+h_{T}^{2})+(\frac{5}{12}g_{1}^{2}+\frac{9}{4}g_{2}^{2}+8g_{3}^{2})h_{B}^{2}+(\frac{5}{4}g_{1}^{2}+\frac{3}{4}g_{2}^{2})h_{E}^{2}\right]\right\}
\end{displaymath}
\begin{displaymath}
+h_{t}\left\{ \frac{5}{2}\left(\frac{1}{4}g_{1}^{2}+\frac{3}{4}g_{2}^{2}\right)h_{N}^{2} +\frac{4721}{648}g_{1}^{4}-\frac{3}{4}g_{1}^{2}g_{2}^{2}+\frac{19}{9}g_{1}^{2}g_{3}^{2}\right\}
\end{displaymath}
\begin{displaymath}
+h_{t}\left\{ -\frac{19}{4}g_{2}^{4}+9g_{2}^{2}g_{3}^{2}-\frac{892}{9}g_{3}^{4}\right\}
\end{displaymath}
\begin{displaymath}
A\beta_{h_{T}}^{(1)}=h_{T}\left[\frac{3}{2}(h_{T}^{2}-h_{B}^{2})+3(h_{t}^{2}+h_{T}^{2}+h_{B}^{2})+h_{N}^{2}+h_{E}^{2}\right]
\end{displaymath}
\begin{displaymath}
+h_{T}\left[ -\frac{17}{12}g_{1}^{2}-\frac{9}{4}g_{2}^{2}-8g_{3}^{2}\right]
\end{displaymath}
\begin{displaymath}
A^{2}\beta_{h_{T}}^{(2)}=h_{T}\left\{\frac{3}{2}h_{T}^{4}-h_{T}^{2}h_{B}^{2}-\frac{1}{4}h_{B}^{2}h_{T}^{2}+\frac{11}{4}h_{B}^{4}\right\}
\end{displaymath}
\begin{displaymath}
+h_{T}\left\{\left(\frac{5}{4}h_{B}^{2}-\frac{9}{4}h_{T}^{2}\right)\left(3(h_{t}^{2}+h_{T}^{2}+h_{B}^{2})+h_{E}^{2}+h_{N}^{2}\right)\right\}
\end{displaymath}
\begin{displaymath}
+h_{T}\left\{-\frac{9}{4}\left(3(h_{t}^{4}+h_{T}^{4}+h_{B}^{4})+h_{E}^{4}+h_{N}^{4}-\frac{2}{3}h_{T}^{2}h_{B}^{2}-\frac{2}{9}h_{E}^{2}h_{N}^{2}\right)\right\}
\end{displaymath}
\begin{displaymath}
+h_{T}\left\{ \frac{\lambda^{2}}{6}-\frac{2 \lambda}{3}(3h_{T}^{2}+h_{B}^{2}) + \left(\frac{223}{48}g_{1}^{2}+\frac{135}{16}g_{2}^{2}+16g_{3}^{2}\right)h_{T}^{2}\right\}
\end{displaymath}
\begin{displaymath}
+h_{T}\left\{ -\left(\frac{43}{48}g_{1}^{2}-\frac{9}{16}g_{2}^{2}+16g_{3}^{2}\right)h_{B}^{2} +\frac{5}{2}\left[ \left(\frac{17}{12}g_{1}^{2}+\frac{9}{4}g_{2}^{2}+8g_{3}^{2}\right)(h_{t}^{2}+h_{T}^{2})\right]\right\}
\end{displaymath}
\begin{displaymath}
+h_{T}\left\{\frac{5}{2}\left[\left(\frac{5}{12}g_{1}^{2}+\frac{9}{4}g_{2}^{2}+8g_{3}^{2}\right)h_{B}^{2}+\left(\frac{5}{4}g_{1}^{2}+\frac{3}{4}g_{2}^{2}\right)h_{E}^{2}+\left(\frac{1}{4}g_{1}^{2}+\frac{3}{4}g_{2}^{2}\right)h_{N}^{2}\right]\right\}
\end{displaymath}
\begin{displaymath}
+h_{T}\left\{ \frac{4721}{648}g_{1}^{4}-\frac{3}{4}g_{1}^{2}g_{2}^{2}+\frac{19}{9}g_{1}^{2}g_{3}^{2}-\frac{19}{4}g_{2}^{4}+9g_{2}^{2}g_{3}^{2}-\frac{892}{9}g_{3}^{4}\right\}
\end{displaymath}
\begin{displaymath}
A\beta_{h_{B}}^{(1)}=h_{B}\left[\frac{3}{2}(h_{B}^{2}-h_{T}^{2})+3(h_{t}^{2}+h_{T}^{2}+h_{B}^{2})+h_{N}^{2}+h_{E}^{2}\right]
\end{displaymath}
\begin{displaymath}
+h_{B}\left[-\frac{5}{12}g_{1}^{2}-\frac{9}{4}g_{2}^{2}-8g_{3}^{2}\right]
\end{displaymath}
\begin{displaymath}
A^{2}\beta_{h_{B}}^{(2)}= h_{B}\left\{\frac{3}{2}h_{B}^{4}-h_{B}^{2}h_{T}^{2}-\frac{1}{4}h_{T}^{2}h_{B}^{2}+\frac{11}{4}h_{T}^{4}+\left(\frac{5}{4}h_{T}^{2}-\frac{9}{4}h_{B}^{2}\right)\left(3(h_{t}^{2}+h_{T}^{2}+h_{B}^{2})+h_{E}^{2}+h_{N}^{2}\right)\right\}
\end{displaymath}
\begin{displaymath}
+h_{B}\left\{ -\frac{9}{4}\left(3(h_{t}^{4}+h_{T}^{4}+h_{B}^{4})+h_{E}^{4}+h_{N}^{4}-\frac{2}{3}h_{T}^{2}h_{B}^{2}-\frac{2}{9}h_{N}^{2}h_{E}^{2}\right)+\frac{\lambda^{2}}{6}\right\}
\end{displaymath}
\begin{displaymath}
+h_{B}\left\{ -\frac{2}{3}\lambda(3h_{B}^{2}+h_{T}^{2})+\left(\frac{187}{48}g_{1}^{2}+\frac{135}{16}g_{2}^{2}+16g_{3}^{2}\right)h_{B}^{2}-\left(\frac{79}{48}g_{1}^{2}-\frac{9}{16}g_{2}^{2}+16g_{3}^{2}\right)h_{T}^{2}\right\}
\end{displaymath}
\begin{displaymath}
+h_{B}\left\{\frac{5}{2}\left[\left(\frac{17}{12}g_{1}^{2}+\frac{9}{4}g_{2}^{2}+8g_{3}^{2}\right)\left(h_{t}^{2}+h_{T}^{2}\right)+\left(\frac{5}{12}g_{1}^{2}+\frac{9}{4}g_{2}^{2}+8g_{3}^{2}\right)h_{B}^{2}+\left(\frac{5}{4}g_{1}^{2}+\frac{3}{4}g_{2}^{2}\right)h_{E}^{2}\right]\right\}
\end{displaymath}
\begin{displaymath}
+h_{B}\left\{ \frac{5}{2}\left(\frac{1}{4}g_{1}^{2}+\frac{3}{4}g_{2}^{2}\right)h_{N}^{2}-\frac{421}{648}g_{1}^{2}-\frac{9}{4}g_{1}^{2}g_{2}^{2}+\frac{31}{9}g_{1}^{2}g_{3}^{2}\right\}
\end{displaymath}
\begin{displaymath}
+h_{B}\left\{ -\frac{19}{4}g_{2}^{4}+9g_{2}^{2}g_{3}^{2}-\frac{892}{9}g_{3}^{4}\right\}
\end{displaymath}
\begin{displaymath}
A\beta_{h_{N}}^{(1)}=h_{N}\left[ \frac{3}{2}\left(h_{N}^{2}-h_{E}^{2}\right)+3(h_{t}^{2}+h_{T}^{2}+h_{B}^{2})+h_{N}^{2}+h_{E}^{2}-\frac{3}{4}g_{!}^{2}-\frac{9}{4}g_{2}^{2}\right]
\end{displaymath}
\begin{displaymath}
A^{2}\beta_{h_{N}}^{(2)}=h_{N}\left\{ \frac{3}{2}h_{N}^{4}-h_{N}^{2}h_{E}^{2}-\frac{1}{4}h_{N}^{2}h_{E}^{2}+\frac{11}{4}h_{E}^{4}\right\}
\end{displaymath}
\begin{displaymath}
+h_{N}\left\{ \left(\frac{5}{4}h_{E}^{2}-\frac{9}{4}h_{N}^{2}\right)\left(3(h_{t}^{2}+h_{T}^{2}+h_{B}^{2})+h_{E}^{2}+h_{N}^{2}\right)\right\}
\end{displaymath}
\begin{displaymath}
+h_{N}\left\{-\frac{9}{4}\left(3(h_{t}^{4}+h_{T}^{4}+h_{B}^{4})+h_{E}^{4}+h_{N}^{4}-\frac{2}{3}h_{T}^{2}h_{B}^{2}-\frac{2}{9}h_{E}^{2}h_{N}^{2}\right)+\frac{\lambda^{2}}{6}\right\}
\end{displaymath}
\begin{displaymath}
+h_{N}\left\{ -\frac{2}{3}\lambda\left(h_{N}^{2}+h_{E}^{2}\right)+\left(\frac{337}{48}g_{1}^{2}+\frac{135}{16}g_{2}^{2}\right)h_{N}^{2}-\left(\frac{207}{48}g_{1}^{2}-\frac{9}{16}g_{2}^{2}\right)h_{E}^{2}\right\}
\end{displaymath}
\begin{displaymath}
+h_{N}\left\{ \frac{5}{2}\left[\left(\frac{17}{12}g_{1}^{2}+\frac{9}{4}g_{2}^{2}+8g_{3}^{2}\right)\left(h_{t}^{2}+h_{T}^{2}\right)+\left(\frac{5}{12}g_{1}^{2}+\frac{9}{4}g_{2}^{2}+8g_{3}^{2}\right)h_{B}^{2}+\left(\frac{5}{4}g_{1}^{2}+\frac{3}{4}g_{2}^{2}\right)h_{E}^{2}\right]\right\}
\end{displaymath}
\begin{displaymath}
+h_{N}\left\{ \frac{5}{2}\left(\frac{1}{4}g_{1}^{2}+\frac{3}{4}g_{2}^{2}\right)h_{N}^{2}+\frac{811}{72}g_{1}^{4}+\frac{9}{20}g_{1}^{2}g_{2}^{2}-\frac{19}{4}g_{2}^{4}\right\}
\end{displaymath}
\begin{displaymath}
A\beta_{h_{E}}^{(1)}=h_{E}\left[\frac{3}{2}(h_{E}^{2}-h_{N}^{2})+3(h_{t}^{2}+h_{T}^{2}+h_{B}^{2})+h_{N}^{2}+h_{E}^{2} -\frac{15}{4}g_{1}^{2}-\frac{9}{4}g_{2}^{2}\right]
\end{displaymath}
\begin{displaymath}
A^2\beta_{h_{E}}^{(2)}=h_{E}\left\{ \frac{3}{2}h_{E}^{4}-h_{N}^{2}h_{E}^{2}-\frac{1}{4}h_{E}^{2}h_{N}^{2}+\frac{11}{4}h_{N}^{4}\right\}
\end{displaymath}
\begin{displaymath}
+h_{E}\left\{ \left(\frac{5}{4}h_{N}^{2}-\frac{9}{4}h_{E}^{2}\right)\left(3(h_{t}^{2}+h_{T}^{2}+h_{B}^{2})+h_{E}^{2}+h_{N}^{2}\right)\right\}
\end{displaymath}
\begin{displaymath}
+h_{E}\left\{-\frac{9}{4}\left(3(h_{t}^{4}+h_{T}^{4}+h_{B}^{4})+h_{E}^{4}+h_{N}^{4}-\frac{2}{3}h_{T}^{2}h_{B}^{2}-\frac{2}{9}h_{E}^{2}h_{N}^{2}\right)+\frac{\lambda^{2}}{6} \right\}
\end{displaymath}
\begin{displaymath}
+h_{E}\left\{-\frac{2}{3}\lambda\left(3h_{E}^{2}+h_{N}^{2}\right)+\left(\frac{387}{48}g_{1}^{2}+\frac{135}{16}g_{2}^{2}\right)h_{E}^{2}-\left(\frac{269}{48}g_{1}^{2}-\frac{9}{16}g_{2}^{2}\right)h_{N}^{2}\right\}
\end{displaymath}
\begin{displaymath}
+h_{E}\left\{\frac{5}{2}\left[\left(\frac{17}{12}g_{1}^{2}+\frac{9}{4}g_{2}^{2}+8g_{3}^{2}\right)\left(h_{t}^{2}+h_{T}^{2}\right)+\left(\frac{5}{12}g_{1}^{2}+\frac{9}{4}g_{2}^{2}+8g_{3}^{2}\right)h_{B}^{2}+\left(\frac{5}{4}g_{1}^{2}+\frac{3}{4}g_{2}^{2}\right)h_{E}^{2}\right]\right\}
\end{displaymath}
\begin{displaymath}
+h_{E}\left\{ \frac{5}{2}\left(\frac{1}{4}g_{1}^{2}+\frac{3}{4}g_{2}^{2}\right)h_{N}^{2}+\frac{1811}{72}g_{1}^{4}+\frac{9}{4}g_{1}^{2}g_{2}^{2}-\frac{19}{4}g_{2}^{4}\right\}
\end{displaymath}
\begin{displaymath}
A\beta_{g_{3}}^{(1)}=-\frac{17}{3}g_{3}^{3}
\end{displaymath}
\begin{displaymath}
A^{2}\beta_{g_{3}}^{(2)}=g_{3}^{3}\left(-2(h_{t}^{2}+h_{T}^{2}+h_{B}^{2})-\frac{2}{3}g_{3}^{2}+6g_{2}^{2}+\frac{22}{9}g_{1}^{2}\right)
\end{displaymath}
\begin{displaymath}
A\beta_{g_{2}}^{(1)}=-\frac{11}{6}g_{2}^{3}
\end{displaymath}
\begin{displaymath}
A^{2}\beta_{g_{2}}^{(2)}=g_{2}^{3}\left(-\frac{3}{2}(h_{t}^{2}+h_{T}^{2}+h_{B}^{2})-\frac{1}{2}(h_{E}^{2}+h_{N}^{2})+16g_{3}^{2}+\frac{133}{6}g_{2}^{2}+\frac{55}{30}g_{1}^{2}\right)
\end{displaymath}
\begin{displaymath}
A\beta_{g_{1}}^{(1)}=\frac{163}{18}g_{1}^{3}
\end{displaymath}
\begin{displaymath}
A^{2}\beta_{g_{!}}^{(2)}=g_{!}^{3}\left(-\frac{17}{6}h_{t}^{2}-\frac{17}{6}h_{T}^{2}-\frac{5}{6}h_{B}^{2}-\frac{5}{2}h_{E}^{2}-\frac{1}{2}h_{N}^{2}\right)
\end{displaymath}
\begin{displaymath}
+g_{1}^{3}\left(\frac{176}{9}g_{3}^{2}+\frac{11}{2}g_{2}^{2}+\frac{787}{54}g_{1}^{2}\right)
\end{displaymath}
\begin{displaymath}
A\beta_{m^{2}}^{(1)}=m^{2}\left(2\lambda+6(h_{t}^{2}+h_{T}^{2}+h_{B}^{2})+2(h_{E}^{2}+h_{N}^{2})-\frac{9}{2}g_{2}^{2}-\frac{3}{2}g_{1}^{2}\right)
\end{displaymath}
\begin{displaymath}
A\gamma_{\phi}^{(1)}=3\left(h_{t}^{2}+h_{T}^{2}+h_{B}^{2}\right)+h_{E}^{2}+h_{N}^{2}-\frac{9}{4}g_{2}^{2}-\frac{3}{4}g_{1}^{2}
\end{displaymath}
\begin{displaymath}
A^{2}\gamma_{\phi}^{(2)}=\frac{1}{6}\lambda+\frac{5}{2}\left\{\left(\frac{17}{12}g_{1}^{2}+\frac{9}{4}g_{2}^{2}+8g_{3}^{2}\right)\left(h_{t}^{2}+h_{T}^{2}\right)+\left(\frac{5}{12}g_{1}^{2}+\frac{9}{4}g_{2}^{2}+8g_{3}^{2}\right)h_{B}^{2}\right\}
\end{displaymath}
\begin{displaymath}
+\frac{5}{2}\left\{\left(\frac{5}{4}g_{1}^{2}+\frac{3}{4}g_{2}^{2}\right)h_{E}^{2}+\left(\frac{1}{4}g_{1}^{2}+\frac{3}{4}g_{2}^{2}\right)h_{N}^{2}\right\}
\end{displaymath}
\begin{displaymath}
-\frac{9}{4}\left\{3(h_{t}^{4}+h_{T}^{4}+h_{B}^{4})+h_{E}^{4}+h_{N}^{4}-\frac{2}{3}h_{T}^{2}h_{B}^{2}-\frac{2}{9}h_{E}^{2}h_{N}^{2}\right\}
\end{displaymath}
\begin{displaymath}
+\frac{9}{16}g_{1}^{2}g_{2}^{2}+\frac{1693}{288}g_{1}^{4}-\frac{191}{32}g_{2}^{4}
\end{displaymath}
\subsection{MSSM4}
\begin{displaymath}
\frac{d g_{1}}{dt}= \frac{g_{1}^{3}}{(4 \pi)^{2}}\left[ \frac{43}{5}+\frac{1}{\left(4\pi\right)^{2}}\left(\frac{787}{75}g_{1}^{2}+\frac{33}{5}g_{2}^{2}+\frac{352}{15}g_{3}^{2}-\frac{18}{15}h_{E}^{2}-\frac{14}{5}h_{B}^{2}-\frac{26}{5}(h_{t}^{2}+h_{T}^{2})\right)\right]
\end{displaymath}
\begin{displaymath}
\frac{d g_{2}}{dt}=\frac{g_{2}^{3}}{(4\pi)^{2}}\left[3 +\frac{1}{(4\pi)^{2}}\left( \frac{11}{5}g_{1}^{2}+39g_{2}^{2}+32g_{3}^{2}-2h_{E}^{2}-6(h_{B}^{2}+h_{t}^{2}+h_{T}^{2})\right)\right]
\end{displaymath}
\begin{displaymath}
\frac{d g_{3}}{d t}=\frac{g_{3}^{3}}{(4 \pi)^{2}}\left[ -1 +\frac{1}{(4\pi)^{2}}\left(\frac{44}{15}g_{1}^{2}+12g_{2}^{2}+\frac{110}{3}g_{3}^{2}-4(h_{B}^{2}+h_{T}^{2}+h_{t}^{2})\right)\right]
\end{displaymath}
\begin{displaymath}
\frac{d h_{E}}{dt}=\frac{h_{E}}{16\pi^{2}}\left(-\frac{9}{5}g_{1}^{2}-3g_{2}^{2}+3h_{B}^{2}+4h_{E}^{2}+h_{N}^{2}\right)
\end{displaymath}
\begin{displaymath}
+\frac{h_{E}}{(16\pi^{2})^{2}}\left(\frac{171}{10}g_{1}^{4}+\frac{27}{2}g_{2}^{4}+\frac{9}{5}g_{1}^{2}g_{2}^{2}+16h_{B}^{2}g_{3}^{2}-\frac{2}{5}h_{B}^{2}g_{1}^{2}+\frac{6}{5}h_{E}^{2}g_{1}^{2}+6h_{E}^{2}g_{2}^{2}\right)
\end{displaymath}
\begin{displaymath}
+\frac{h_{E}}{(16\pi^{2})^{2}}\left(-9h_{B}^{4}-10h_{E}^{4}-3h_{B}^{2}h_{T}^{2}-h_{E}^{2}h_{N}^{2}-9h_{E}^{2}h_{B}^{2}-3h_{N}^{4}-3h_{N}^{2}h_{t}^{2}-3h_{N}^{2}h_{t}^{2}-2h_{E}^{2}h_{N}^{2}\right)
\end{displaymath}
\begin{displaymath}
\frac{d h_{B}}{dt}=\frac{h_{B}}{16\pi^{2}}\left(-\frac{7}{15}g_{1}^{2}-3g_{2}^{2}-\frac{16}{3}g_{3}^{2}+6h_{B}^{2}+h_{T}^{2}+h_{E}^{2}\right)
\end{displaymath}
\begin{displaymath}
+\frac{h_{B}}{(16\pi^{2})^{2}}\left(\frac{371}{90}g_{1}^{4}+\frac{27}{2}g_{2}^{4}+\frac{80}{9}g_{3}^{4}+g_{1}^{2}g_{2}^{2}+\frac{8}{9}g_{1}^{2}g_{3}^{2}+8g_{2}^{2}g_{3}^{2}\right)
\end{displaymath}
\begin{displaymath}
+\frac{h_{B}}{(16\pi^{2})^{2}}\left( 16 h_{B}^{2}g_{3}^{2}+\frac{2}{5}g_{1}^{2}h_{B}^{2}+\frac{4}{5}h_{T}^{2}g_{1}^{2}+\frac{6}{5}h_{E}^{2}g_{1}^{2}+6h_{B}^{2}g_{2}^{2}-22h_{B}^{4}-3h_{E}^{4}\right)
\end{displaymath}
\begin{displaymath}
+\frac{h_{B}}{(16\pi^{2})^{2}}\left(-5h_{B}^{2}h_{T}^{2}-h_{E}^{2}h_{N}^{2}-5h_{T}^{4}-3h_{t}^{2}h_{T}^{2}-3h_{B}^{2}h_{E}^{2}-h_{N}^{2}h_{E}^{2}\right)
\end{displaymath}
\begin{displaymath}
\frac{d h_{N}}{dt}= \frac{h_{N}}{(16\pi^{2})}\left(-\frac{3}{5}g_{1}^{2}-3g_{2}^{2}+3h_{T}^{2}+3h_{t}^{2}+4h_{N}^{2}+h_{E}^{2}\right)
\end{displaymath}
\begin{displaymath}
+\frac{h_{N}}{(16\pi^{2})^{2}}\left( \frac{267}{50}g_{1}^{4}+\frac{27}{2}g_{2}^{4}+\frac{9}{5}g_{!}^{2}g_{2}^{2}+16h_{t}^{2}g_{3}^{2}+16h_{T}^{2}g_{3}^{2}+\frac{6}{5}h_{E}^{2}g_{1}^{2}+6h_{N}^{2}g_{2}^{2}\right)
\end{displaymath}
\begin{displaymath}
+\frac{h_{N}}{(16 \pi^{2})^{2}}\left( \frac{6}{5}h_{N}^{2}g_{1}^{2}+\frac{4}{5}h_{t}^{2}g_{1}^{2}+\frac{4}{5}h_{T}^{2}g_{1}^{2}-9h_{t}^{4}-9h_{T}^{4}-10h_{N}^{4}\right)
\end{displaymath}
\begin{displaymath}
+\frac{h_{N}}{(16\pi^{2})^{2}}\left( -3h_{B}^{2}h_{T}^{2}-3h_{E}^{2}h_{N}^{2}-3h_{E}^{4}-3h_{E}^{2}h_{B}^{2}-9h_{N}^{2}h_{t}^{2}-9h_{E}^{2}h_{T}^{2}\right)
\end{displaymath}
\begin{displaymath}
\frac{d h_{t}}{dt}=\frac{h_{t}}{16 \pi^{2}}\left( -\frac{13}{15}g_{1}^{2}-3g_{2}^{2}-\frac{16}{3}g_{3}^{2}+6h_{t}^{2}+h_{N}^{2}\right)
\end{displaymath}
\begin{displaymath}
+\frac{h_{t}}{(16\pi^{2})^{2}}\left(\frac{3523}{450}g_{1}^{4}+\frac{27}{2}g_{2}^{4}+\frac{80}{9}g_{3}^{4}+g_{1}^{2}g_{2}^{2}+\frac{136}{45}g_{1}^{2}g_{3}^{2}+8g_{2}^{2}g_{3}^{2}+16h_{t}^{2}g_{3}^{2}\right)
\end{displaymath}
\begin{displaymath}
+\frac{h_{t}}{(16\pi^{2})^{2}}\left(\frac{6}{5}h_{t}^{2}g_{1}^{2}+6h_{t}^{2}g_{2}^{2}-22h_{t}^{4}-3h_{N}^{4}-3h_{B}^{2}h_{T}^{2}-h_{E}^{2}h_{N}^{2}\right)
\end{displaymath}
\begin{displaymath}
+\frac{h_{t}}{(16\pi^{2})^{2}}\left(-9h_{t}^{2}h_{T}^{2}-3h_{t}^{2}h_{N}^{2}\right)
\end{displaymath}
\begin{displaymath}
\frac{d h_{T}}{dt}=\frac{h_{T}}{16 \pi^{2}}\left(-\frac{13}{15}g_{1}^{2}-3g_{2}^{2}-\frac{16}{3}g_{3}^{2}+6h_{T}^{2}+3h_{t}^{2}+h_{N}^{2}\right)
\end{displaymath}
\begin{displaymath}
+\frac{h_{T}}{(16 \pi^{2})^{2}}\left(\frac{3523}{450}g_{1}^{4}+\frac{27}{2}g_{2}^{4}+\frac{80}{9}g_{3}^{4}+g_{1}^{2}g_{2}^{2}+\frac{136}{45}g_{1}^{2}g_{3}^{2}+8g_{2}^{2}g_{3}^{2}+16h_{T}^{2}g_{3}^{2}\right)
\end{displaymath}
\begin{displaymath}
+\frac{h_{T}}{(16\pi^{2})^{2}}\left(16h_{t}^{2}g_{3}^{2}+\frac{6}{5}h_{T}^{2}g_{1}^{2}+\frac{4}{5}h_{t}^{2}g_{1}^{2}+6h_{T}^{2}g_{2}^{2}-22h_{T}^{4}-9h_{t}^{4}-3h_{N}^{4}\right)
\end{displaymath}
\begin{displaymath}
+\frac{h_{T}}{(16\pi^{2})^{2}}\left(-5h_{B}^{2}h_{T}^{2}-h_{T}^{2}h_{E}^{2}-5h_{B}^{4}-9h_{T}^{2}h_{t}^{2}-3h_{T}^{2}h_{N}^{2}\right)
\end{displaymath}
\subsection{$\Delta \Pi$}
In this appendix we present the expression for $\Delta \Pi$:
\begin{displaymath}
\Delta \Pi = \Delta \Pi_{ff}+\Delta\Pi_{bosons}+\Delta\Pi_{scalar}
\end{displaymath}
(i) Fermion contribution
\begin{displaymath}
\Delta\Pi_{ff}=\sum_{i=1}^{5} \frac{c_{i}h_{i}^{2}}{8\pi^{2}}\left\{ -2M_{i}^{2}\left[Z\left(\frac{M_{i}^{2}}{M_{H}^{2}}\right)-2\right]+\frac{1}{2}M_{H}^{2}\left[\log\frac{M_{i}^{2}}{\mu^{2}}+Z\left(\frac{M_{i}^{2}}{M_{H}^{2}}\right)-2\right]\right\}.
\end{displaymath}
where $c_{i}$ = 1 for i = (N,E) and $c_{i}$ = 3 for i = (t,T,B).

(ii) Gauge boson and Goldstone boson contribution
\begin{displaymath}
\Delta\Pi_{bosons}=\frac{g_{2}^{2}M_{W}^{2}}{8\pi^{2}}\left[-3+\frac{5M_{H}^{2}}{4M_{W}^{2}}+\frac{1}{2}\left(3-\frac{M_{H}^{2}}{M_{W}^{2}}+\frac{M_{H}^{4}}{4M_{W}^{4}}\right)Z\left(\frac{M_{W}^{2}}{M_{H}^{2}}\right)\right.
\end{displaymath}
\begin{displaymath}
\left.
-\frac{M_{H}^{4}}{8M_{W}^{4}}\log\frac{M_{H}^{2}}{M_{W}^{2}}-\frac{3M_{H}^{2}}{4M_{W}^{2}}\log\frac{M_{W}^{2}}{\mu^{2}}\right]+\frac{1}{2}\left\{\begin{array}{c}
M_{W} \rightarrow M_{Z} \\ g_{2}^{2} \rightarrow g_{2}^{2} + g_{1}^{2} 
\end{array} \right\}
\end{displaymath}
(iii) Pure scalar contribution
\begin{displaymath}
\Delta\Pi_{scalar}=\frac{3}{128\pi^{2}}\frac{g_{2}^{2}M_{H}^{4}}{M_{W}^{2}}\left[ \pi\sqrt{3}-8+4\log\frac{M_{H}^{2}}{\mu^{2}(t_{v})}\right]
\end{displaymath}
where the function $Z(x)$ is defined as:
\begin{displaymath}
Z(x)=\left\{ \begin{array}{ll}
     2A\tan^{-1}(\frac{1}{A}) & \mbox{if $x>\frac{1}{4}$} \\
     A\log\left[\frac{1+A}{1-A}\right] & \mbox{if$x<\frac{1}{4}$}
	\end{array}
	\right.
\end{displaymath}
\begin{displaymath}
A \equiv |1-4x|^{\frac{1}{2}}
\end{displaymath}

\begin{figure}
\caption{Plots of the lightest Higgs boson mass $M_{H}$ as a function of $m_{4}$ for $M_{t}$ = 177 GeV; (a) no threshold corrections and $M_{susy}$= 1 TeV and (b) maximum threshold corrections and $M_{susy}$ = 1 TeV. The solid and dotted lines correspond to the MSSM4 and the MSSM3, respectively. The upper and lower lines of each correspond to $\cos^{2}(2 \beta)$= 1 and 0, respectively.}
\label{1}
\end{figure}
\begin{figure}
\caption{Same as figure (1), but with $M_{susy}$= 10 TeV.}
\label{2}
\end{figure}
\begin{figure}
\caption{Plots of the running Higgs mass $m_{h}$ for $M_{susy}$= 1 TeV; (a) $\cos^{2}(2 \beta)$ = 1 and (b) $\cos^{2}(2 \beta)$ = 0. The solid and dotted lines correspond to one-loop and two-loop results, respectively}
\label{3}
\end{figure}
\begin{figure}
\caption{Same as figure (3), but with $M_{susy}$ = 10 TeV.}
\label{4}
\end{figure}
\begin{figure}
\caption{Plots of SM Higgs boson mass lower bound and absolute maximum MSSM Higgs boson mass upper bound for $\Lambda = 10^{19}$ GeV; (a) $M_{susy}=$ 10 TeV and (b) $M_{susy}$ = 1 TeV. The uppermost solid curve is teh SM4 lower bound, while the lower solid curve is the MSSM4 upper bound. The uppermost dotted line is the SM3 lower bound, and the bottom dotted line represents the MSSM3 upper bound. The MSSM(3,4) bounds are calculated with $\cos^{2}(2 \beta)$ = 1 and maximum threshold corrections.}
\label{5}
\end{figure}

\begin{thebibliography}{99}

\bibitem{1} For reviews, see H.P. Nilles, Phys. Rep. 110 (1984) 1; H. E. Haber and G. Kane, Phys. Rep. 117 (1985) 76.
\bibitem{2} J.F. Gunion et al., The Higgs Hunters Guide, Redwood City, Calif.: Addison Wesley, c1990.
\bibitem{3} M. Sher, Phys. Rep. 179 (1989) 273, and references therein.
\bibitem{4} J.A. Casas, J.R. Espinosa, M. Quiros, A. Riotto, Nucl. Phys. B 436 (1995) 3; B 439 (1995) 466 (E).
\bibitem{5} J.A. Casas, J.R. Espinosa, and M.Quiros, Phys. Lett. B 342 (1995) 171; B 382 (1996) 374.
\bibitem{6} J. Kodaira, Y. Yasui, and K. Sasaki, Phys. Rev. D 50 (1994) 50.
\bibitem{7} Y. Okada, M. Yamaguchi, and T. Yanagida, Phys. Lett B 262 (1991) 54; Prog. Theor. Phys. 85 (1991) 1; H.E. Haber and R. Hempfling, Phys. Rev. Lett. 66 (1991) 1815; Phys. Rev. D 48 (1993) 4280; J. Ellis, G. Ridolfi, and F. Zwirner, Phys. Lett. B 257 (1991) 83; R. Barbieri, M. Frigeni, and F. Caravaglios, Phys. Lett. B 258 (1991) 167.
\bibitem{8} G. Altarelli and G. Isidori, Phys. Lett. B 337 (1994) 141.
\bibitem{9} K. Kang and S. K. Kang, preprint hep-ph/9702355, BROWN-HET-1092, preprint hep-ph/9708409; J. Timmerman, talk presented at the International Symposium on lepton-photon interactions, Hamburg, 1997.
\bibitem{10} Particle Data Group, R. M. Barnett et al., Phys. Rev. D 54 (1996) 1; LEP Electroweak Working Group, Internal Note, preprint LEPEWWG/97-02.
\bibitem{11} H.B. Nielsen, A.V. Novikov, and M.S. Vysotsky, Phys. Lett. B 374 (1996) 127.
\bibitem{12} V. Novikov, preprint hep-ph/9606318 (June 1996); LEP1.5 Collaboration, J. Nachtman, in Electroweak Interactions and Unified Theories, Proceedings of the 31st rencontres de Moriond, Les Arcs, France (1996), preprint hep-ex/960615.
\bibitem{13} K.S. Babu and E. Ma, Z. Phys. C 29 (1985) 45.
\bibitem{14} S.K. Kang and G.T. Park, Mod. Phys. Lett. A 12 (1997) 553, preprint hep-ph/9702355.
\bibitem{15} S.K. Kang, Phys. Rev. D 54 (1996) 7077.
\bibitem{16} Y. Okada et al., Ref \cite{7}.
\bibitem{17} C. Ford, D.R.T. Jones, P.W. Stephenson, and M.B. Einhorn, Nucl. Phys. B 395 (1993) 17.
\bibitem{18} J.F. Gunion, D.W. McKay, and H. Pois, Phys. Lett. B 334 (1994) 339; J.F. Gunion, D.W. McKay, and H. Pois, Phys. Rev. D 53 (1996) 53.
\bibitem{19} M. Lindner, Z. Phys. C 31 (1986) 295.
\bibitem{20} M. Machacek and M. Vaughn, Nucl. Phys. B 222 (1983) 83; B 236 (1984) 221; B 249 (1985) 70.
\bibitem{21} J.R. Espinosa and M. Quiros, Phys. Lett. B 353 (1995) 257.
\end{thebibliography}
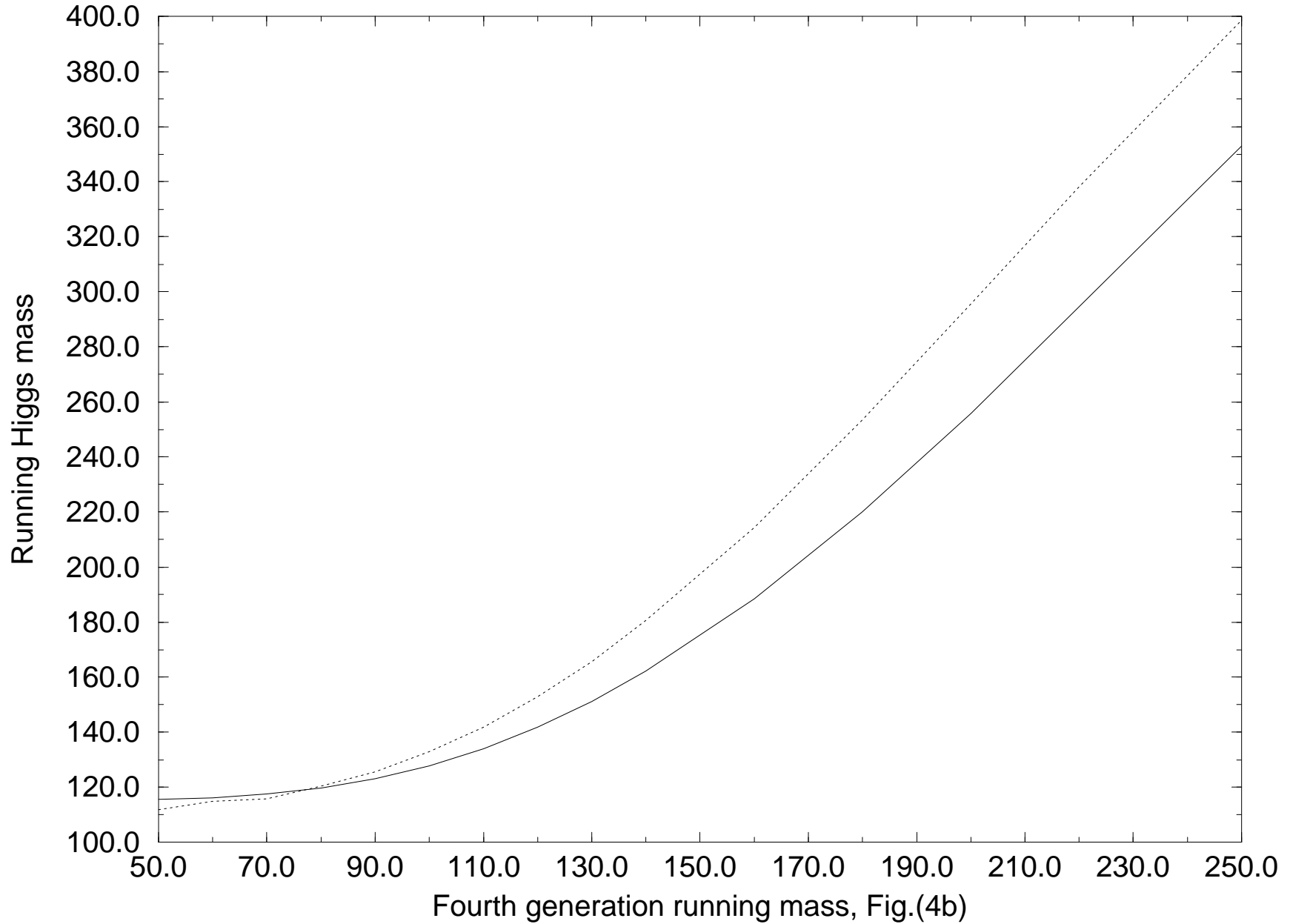
\end{document}